\documentstyle[11pt]{article}                                                  
                                            
\begin{document}                                                                
\begin{center}                                                                  
{\Large Parton sum rules and improved scaling variable}

\vspace{4mm}                                                                    
\renewcommand{\thefootnote}{\fnsymbol{footnote}}                                
{\normalsize Bo-Qiang Ma$^{1}$, %and                                                 
Andreas Sch\"afer$^{2}$}                              
                                                                                
\vspace{4mm}                                                                    
{\small                                                                         
        $^1$CCAST(World Laboratory), P.~O.~Box 8730, 
         Beijing 100080, and 
         Institute of High Energy Physics, Academia Sinica, P.~O.~Box        
         918(4), Beijing 100039, China                                           
%         \footnote{Mailing address}                                             
        }                                                                       
                                                                                
{\small                                                                         
        $^2$Institut f\"ur Theoretische Physik der Universit\"at                
        Frankfurt am Main, Postfach 11 19 32, D-60054 Frankfurt,                
        Germany}

\vspace{4mm}

{\ \large \bf Abstract }                                                        
                                                                                
\end{center}                                                                    
                                                                                
The effect from quark masses and transversal motion on the Gottfried,
Bjorken, and              
Ellis-Jaffe sum rules is examined by using                                  
a quark-parton model     
of nucleon structure functions based on an improved scaling variable. 
Its use results in
corrections to the Gottfried, Bjorken, and Ellis-Jaffe sum rules.
We use the Brodsky-Huang-Lepage prescription of light-cone wavefunctions 
to estimate the size of the corrections. 
We constrain our choice of parameters by the roughly known higher
twist corrections to the Bjorken sum rule and find that the resulting
corrections to the Gottfried and Ellis-Jaffe sum rules are 
relevant, though not large enough to explain the observed 
sum rule violations.
                                                                                
\vspace{8mm}                                                                    
                                                                                
\noindent                                                                       
PACS number(s): 13.60.Hb, 11.50.Li, 11.80.Cr, 12.40.Aa                          
                                                                                
\break                                                                          
%\renewcommand{\theequation}{\thesection.\arabic{equation}}                     
%\renewcommand{\ref}{(\ref{theequation})}                                       

%\renewcommand{\thesection}{\Roman{section}.}                                   
%\section{INTRODUCTION}                                                         
%\renewcommand{\thesection}{\arabic{section}}                                   
                                                                                
Parton sum rules provide information on the quark distributions                 
in nucleons and thus allow for sensitive tests of QCD.
Recently, the Gottfried sum rule (GSR)              
\cite{GSR}                                                                      
violation reported by the New Muon Collaboration                                
(NMC) \cite{NMC91} has inspired a number of discussions on 
flavor dependence of sea distributions in the nucleons                       
\cite{Ma93}.          
For polarized structure functions                                                          
the violation of the Ellis-Jaffe sum rule (EJSR) \cite{EJSR}  
observed at CERN \cite{EMC,SMC} and SLAC \cite{E142,E143}                                   
has triggered extensive investigations \cite{Ans94}.                                      
In this letter we examine the effect of quark masses and                      
transversal motion on structure functions using an improved             
quark-parton model \cite{Ma86,Ma91}                                                  
formulated in the framework of light-cone quantum field theory \cite{LC}.                     
It will be shown that the kinematical corrections to the Gottfried,
Bjorken, and Ellis-Jaffe sum rules are non-trivial.
                                                                                
We first examine the Gottfried sum rule.                                       
According to the quark-parton model, the nucleon structure functions            
$F^N_2(\nu,Q^{2})$  scale in the Bjorken scaling variable                   
$x_B=Q^2/2M\nu$ in the Bjorken limit $\nu \rightarrow \infty$,                  
$Q^2 \rightarrow \infty$ with $Q^2/2M\nu$ fixed, i.e.,                          
\begin{equation}                                                                
F^N_2(\nu,Q^{2})=F_2^N(x_B)=\sum_i e_i^2 x_B[                                   
q_i^N(x_B)+\overline{q}_i^N(x_B)],                                              
\label{eq:nqpm}                                                                 
\end{equation}                                                                  
where $q_i(x_B)$ ($\overline{q}_i(x_B)$) is the quark (anti-quark)              
momentum distribution function, $e_i$ is the charge of of a quark of            
flavor $i$, and $N$ represents the proton $p$ or neutron $n$.                      
The QCD corrections to flavor number                     
conservation  
$\int_0^1 [u^p(x_B)-\overline{u}^p(x_B)]\;{\rm d} x_B=2$ etc. 
are small \cite{Ros79}. Thus the Gottfried sum  
can be expressed as                     
\begin{equation}                                                                
S_{G}=\int_{0}^{1}[F_{2}^{p}(x_B)-F_{2}^{n}(x_B)]\;{\rm d} x_B/x_B ~
=\frac{1}{3}+\int_{0}^{1}\sum_{i} [2\overline{q}_{i}^{p}(x_B)              
 -2\overline{q}_{i}^{n}(x_B)]\;{\rm d} x_B.                                             
\label{eq:ngsp}                                                                 
\end{equation}                                                                  
Under the assumptions of isospin symmetry between                               
proton and neutron, and flavor                               
symmetry in the sea, one arrives at the Gottfried sum              
rule (GSR), $S_{G}=1/3$.
In the NMC experiment, the value of $S_{G}$ was                                 
determined from
the ratio $F_{2}^{d}/F_{2}^{p}$ in                 
the kinematic range of                        
$x=0.004-0.8$ for $Q^{2}=4$ GeV$^{2}$.  Assuming a smooth                        
extrapolation of the data for $F_{2}^{n}/F_{2}^{p}$ from $x=0.8$ to                 
$x=1$, adopting a Regge behavior $ax^{b}$ for                                      
$F_{2}^{p}-F_{2}^{n}$ (a flavor nonsinglet quantity) in the region                       
$x=0.004-0.15$ and then extrapolating it to $x=0$, the NMC reported                                                              
\begin{equation}                                                                
S_{G}=0.235\pm0.026,                                                            
\end{equation}                                                                  
which is significantly smaller than the simple quark-parton-model result of            
1/3.                                                                            
Several different explanations for the                                          
origin of the GSR violation have been proposed,                                 
such as flavor asymmetry of the nucleon sea \cite{Pre91,Pi},                  
isospin symmetry breaking between the proton                                    
and the neutron \cite{Ma92},                                                    
non-Regge behaviors at small $x$ \cite{Mar90},                                  
and nuclear                                                                     
effects like mesonic exchanges in the deuteron \cite{Kap91}                 
and nuclear bindings \cite{Epe92}.  Recently, it has been                   
concluded in \cite{Saw93} by using an improved scaling variable
that the kinematic corrections may account for a significant part               
of the GSR                                                                      
violation. It is the last proposition that we want to reanalyse. 

Several years ago, an improved quark-parton model for nucleon                   
structure functions was developed                                               
\cite{Ma86}                                                                     
based on light-cone                                                               
quantum field theory\cite{LC}.
The resulting nucleon structure functions have                        
the form                                                                        
\begin{equation}                                                                
F^N_2(\nu,Q^{2})=\sum_i e_i^2 x_p 
{\cal K} %q^-/k'^-\;
[q_i^N(x_p)+\overline{q}_i^N(x_p)],                                              
\label{eq:iqpm}                                                                 
\end{equation}                                                                  
where                                                                           
\begin{equation}                                                                
{\cal K}=q^-/k'^-=2M\nu x_p/(m^2+({\bf k}_\perp+{\bf q}_\perp)^2),                       
\label{eq:factor}                                                               
\end{equation}                                                                  
\begin{equation}                                                                
x_p=(A-B)/2(M^2+2M\nu)                                                          
\end{equation}                                                                  
with                                                                            
\begin{equation}                                                                
\begin{array}{clcr}                                                             
A=M^2+2M\nu+({\bf k}_\perp+{\bf q}_\perp)^2+m^2-{\bf k}_\perp ^2                
 -\lambda^2;\\                                                                   
B=(A^2-4[({\bf k}_\perp+{\bf q}_\perp)^2+m^2](M^2+2M\nu))^{1/2},                
\label{eq:xp2}                                                                  
\end{array}                                                                     
\end{equation}                                                                  
$m$ and $\lambda$ are the masses of the struck quark and the                    
spectator treated as a single particle, ${\bf k}_\perp$ is the     
transversal quark momentum, ${\bf q}_\perp$ is the transversal                  
component of the lepton momentum transfer specified by                          
$q_\nu=(q^+,q^-,{\bf q}_\perp)=(0,2\nu,{\bf q}_\perp)$ with                     
$q^2=-Q^2$, and $x_p$ is the improved scaling variable. 
It gives power-law                 
type corrections to Bjorken scaling violation 
which might be present in some experimental data
\cite{Ma86}.   
Let us note that $x_p$ reduces to the           
Bloom-Gilman variable, the Weizmann variable, and the Nachtmann variable
in appropriate approximations. 
The kinematical factor ${\cal K}=q^-/k'^-$ is almost equal 
to unity in the whole           
$Q^2$ region, even when $Q^2$ is small.                                         
Unfortunately, a good use of the improved scaling variable                      
requires more detailed information on the nucleon wave functions 
than presently available.
In this                                                                         
work, we simply
adopt the Brodsky-Huang-Lepage (BHL) prescription \cite{LC}
for the momentum space wave function in a light-cone
SU(6) quark-spectator model for deep-inelastic
scattering \cite{DQ,DQ2} to analyse the effect due to finite
quark masses and transversal motion in the Gottfried sum. 

In the Bjorken limit 
the factor                     
${\cal K}$ reduces to unity and the improved scaling variable                   
$x_p$ becomes identical  to the Bjorken variable
(Eq.~(\ref{eq:iqpm})                                  
reduces to Eq.~(\ref{eq:nqpm})).             
However, at finite $Q^2$ the correct condition for flavor   
number conservation is
$                                   
\int_0^1 [u^p(x_p)-\overline{u}^p(x_p)]\;{\rm d} x_p=2$ etc., and 
Eq.~(\ref{eq:ngsp}) does not hold exactly. Under the                
assumption of a flavor and isospin symmetric sea in the                             
nucleons, we obtain                                                             
\begin{equation}                                                                
%\begin{array}{clcr}            
S_{G}=\int_{0}^{1}(F_{2}^{p}(x_B)-F_{2}^{n}(x_B))\;{\rm d} x_B/x_B
%\\
=                    
\frac{1}{3}\int_{0}^{1} x_p 
{\cal K} %q^-/k'^-\; 
[u_v(x_p)-d_v(x_p)]\;{\rm d} x_B/x_B.            
\label{eq:isgp}                                                                 
%\end{array}
\end{equation}                                                                  

The valence quark distributions $u_v(x)$ and $d_v(x)$ in the 
SU(6) quark-spectator model are expressed by
\begin{equation}
\begin{array}{clcr}
u_{v}(x)=\frac{1}{2}a_S(x)+\frac{1}{6}a_V(x);\\
d_{v}(x)=\frac{1}{3}a_V(x),
\label{eq:ud}
\end{array}
\end{equation}
where $a_D(x)$ ($D=S$ or $V$ for scalar or vector spectators) 
denotes the amplitude for the quark
$q$ is scattered while the spectator is in the diquark state $D$
and is normalized such
that $\int_0^1 {\rm d} x a_D(x)=3$
\cite{DQ,DQ2}.
Thus we can write
%written as
\begin{equation}
a_D(x_p)
\propto \int [{\rm d} x] 
[{\rm d}^2 {\bf k}_{\perp}] |\varphi_{D}(x,{\bf k}_{\perp})|^2
\delta(x-x_p),
\end{equation}
where $\varphi_{D}(x,{\bf k}_{\perp})$ is 
the BHL
light-cone momentum space wave function of the 
quark-spectator 
\begin{equation}
\varphi_{D}(x,{\bf k}_{\perp})
=A_{D} exp\{-\frac{1}{8\beta^2_{D}}[\frac{m^2_q+{\bf k}^2_{\perp}}{x}
+\frac{m^2_D+{\bf k}^2_{\perp}}{1-x}]\}.
\label{eq:BHL}
\end{equation}
Here ${\bf k}_{\perp}$ is the internal quark transversal momentum,
$m_q$ and $m_D$ are the masses for the quark $q$ and spectator $D$,
and $\beta_D$ is the harmonic oscillator scale parameter.
Combining Eqs.~(\ref{eq:isgp}) and (\ref{eq:ud}), we have
\begin{equation}                                                                
S_{G}=
\frac{1}{3}\int_{0}^{1} x_p 
{\cal K} %q^-/k'^-\; 
[\frac{1}{2}a_S(x_p)-\frac{1}{6}a_V(x_p)]\;{\rm d} x_B/x_B            
=\frac{1}{6}<a_S>-\frac{1}{18}<a_V>,
\label{eq:SumG}                                                                 
\end{equation}
where
\begin{equation}
<a_D>=\int_{0}^{1} [{\rm d} x_B] [{\rm d}^2 {\bf k}_\perp]
\frac{x_p}{x_B} {\cal K} \;  
a_D(x_p).            
\end{equation}
%for the theoretical expression of the Gottfried sum 
%measured in the NMC experiments.              

The values of the parameters $\beta_D$, $m_q$ and $m_D$ 
can be adjusted by fitting the hadron properties
such as the electromagnetic form factors, 
the mean charge radiuses, and the
weak decay constants etc. in the relativistic light-cone
quark model \cite{LCQM}. 
We used various sets of parameters allowed by these constraints and
calculated the resulting corrections to Bjorken and Ellis-Jaffe sum
rules for 
various values of $Q^2\gg 1~{\rm GeV}^2$. 
As both mass and transverse momentum 
corrections are higher twist effects the resulting $Q^2$ dependence can be 
fitted for large $Q^2$ by a term $c/Q^2$. 
Sum rule calculations suggest $-0.02\leq c(B)\leq 0.03$ GeV$^2$ 
and  $-0.04\leq c(EJ)\leq 0.01$ GeV$^2$ \cite{Stein}. 
In view of principal uncertainties in the sum rule
approach these numbers are chosen very conservatively and should be 
interpreted as 
constraints. For many otherwise acceptable 
parametrizations our model results in much larger values for
$c$. We used therefore the 
sum rule values to restrict 
the parameter range of our model much tighter. And for this restricted 
parameter 
range the resulting corrections to the polarized sum rules 
were calculated for 
the experimentally relevant small values of $Q^2$ and the corrections to 
the Gottfried sum rule were estimated. All corrections turned out to 
be  noticeable but not large
and the remaining allowed parameter variations have little effect.
We shall present results as example for   
$m_q=220$~MeV,
$\beta_S=\beta_V=220$~MeV, $m_S=400$ MeV, and $m_V=600$ MeV (set I).
(The masses of the scalar and vector spectators should
be different taking into account the spin force from color magnetism
or alternatively from instantons \cite{Web94}.) 
To explore the maximal range of parameters we shall also 
allow for SU(6) asymmetric wavefunctions by choosing the parameters
$m_q=220$~MeV,
$\beta_S=280$~MeV, $\beta_V=180$~MeV, 
$m_S=400$ MeV, and $m_V=600$ MeV (set II). This fit leads to higher twist
corrections which are rather too large to be acceptable, namely $c(B)\simeq
c(EJ)\simeq -0.05$. Still we include it for comparison. 

With the above parameters we find 
for set I: \\
$S_G=0.304$ for $Q^2=3~{\rm GeV}^2$ and
$S_G=0.324$ for $Q^2=10~{\rm GeV}^2$ \\
implying that the correction to Gottfried sum rule is small.
For the `exotic' set II we get: \\
$S_G=0.282$ for $Q^2=3~{\rm GeV}^2$ and
$S_G=0.316$ for $Q^2=10~{\rm GeV}^2$ \\
showing that even with extreme assumptions the kinematic corrections 
can account at most for part of the observed sum rule violation.

There has been a                                                                 
similar work by                                                               
Sawicki and Vary \cite{Saw93}                                                   
on off-shell corrections to the                                             
parton model. They arrived at the conclusion that the              
kinematic corrections may account for a substantial part of the GSR                         
violation. There are some important differences between their and 
our calculation apart from the fact that we use the known bounds 
on higher twist
contributions to constrain the range of allowed parameters.                                                                       
First, they assumed the conservation of four-momentum at the photon-parton      
vertex and used                            
$\tilde{x}                                                                      
=x_B (1+\frac{m^2+2{\bf k}_{\perp} \cdot {\bf q}_{\perp}-k^2}{Q^2})$
as the revised scaling variable.                  
However, in light-cone quantum                                              
field theory \cite{LC}                                                          
``energy'' is not conserved at the photon-parton                            
vertex but only between the                 
initial and final states. The            
scaling variable $x_p$                                                          
we used is therefore different from $\tilde{x}$.                                          
Second, in the NMC experiment, the Gottfried sum was obtained by using          
$S_{G}=\int_{0}^{1}(F_{2}^{p}(x_B)-F_{2}^{n}(x_B))\;{\rm d} x_B/x_B$, where            
the measured data $F_2^N(x_B)$ were divided by the Bjorken                       
scaling variable $x_B$ rather than the revised scaling variable                 
$x_p$ or $\tilde{x}$ in the integration over $x_B$.                             
The use of the improved scaling variables $x_p$ or $\tilde{x}$                  
exhibits a shift of the actual parton                                           
distributions towards higher values of $x_p$ or $\tilde{x}$.                    
While the contribution due to the improved scaling variable                           
$x_p$ tends to reduce the Gottfried sum $S_G$ 
this effect is partially canceled by the factor $1/x_B$ (rather than $1/x_p$).
This aspect was ignored in \cite{Saw93}.                                        
Our kinematic factor ${\cal K}$ is also                                   
larger than the factor $x_B/\tilde{x}$ in \cite{Saw93}
and finally the size of our corrections is constrained by fitting the
higher twist contributions to Bjorken sum rule.                         

We now turn our attention to the Ellis-Jaffe sum rule.
Following Ref.~\cite{Ma86}, 
we obtain for the antisymmetric part of the hadron
scattering tensor                                    
$W^A_{\mu\nu}$ in light-cone quantum field theory                  
\cite{LC}                                                     
\begin{equation}                                                                
W^A_{\mu\nu}=\sum_{q}\int\frac{{\rm d}^2{\bf k}_{\perp}{\rm d} k^+}
{16\pi^3 k^+}             
\frac{\rho_q(\underline{k})}{x} w_{\mu\nu}^A(k,k'),                             
\label{eq:at2}                                                                  
\end{equation}                                                                  
\begin{equation}
w_{\mu\nu}^A(k,k')=i\epsilon_{\mu\nu\lambda\sigma}e_q^2 q^{\lambda}
s^{\sigma} \delta(p^-+q^--k'^--\sum_{i=2}^{n}k_i^-)/k'^+,
%\label{eq:atq}
\end{equation} 
where $\rho_q(\underline{k})$ is the distribution function for the           
quark $q$ in the nucleon bound state as a function of the 
light-cone three-momentum
($\underline{k}=(k^+,{\bf k}_{\perp})$). For $g_1$ this implies
\begin{equation}                                                                
g_1(\nu,Q^2)S^{\sigma}=\sum_{q} e^2_q p\cdot q                                  
\int\frac{{\rm d}^2{\bf k}_{\perp}{\rm d} k^+}{16\pi^3 k^+}                                  
\frac{\rho_q(\underline{k})}{x}                                                 
s^{\sigma} \delta(p^-+q^--k'^--\sum_{i=2}^{n}k_i^-)/k'^+.                       
\label{eq:g11}                                                                  
\end{equation}                                                                  
Calculating the $+$ component of Eq.~(\ref{eq:g11}) and                         
treating the $\delta$-function as in Ref.~\cite{Ma86}, we obtain                
$(\Delta q(x)=q^{\uparrow}(x)-q^{\downarrow}(x))$
\begin{equation}                                                                
g_1(\nu,Q^2)=\frac{1}{2}\sum_{q} e^2_q                                          
\int_0^1 
{\cal K} %q^-/k'^-\; 
\Delta q(x) \delta(x-x_p)\;{\rm d} x,                              
%=\frac{1}{18}[4\Delta u(x_p)+\Delta d(x_p)]                                     
\label{eq:g12}                                                                  
\end{equation}                                                                  
with $q^{\uparrow}(x)$ ($q^{\downarrow}(x)$) being the probability              
of finding a quark of flavor $q$ with light-cone helicity parallel              
(antiparallel) to the target spin \cite{Ma91}.                                 
%Thus the parton interpretation of $g_1$ reads:
%\begin{equation}                                                                
%g^N_1(\nu,Q^{2})                                                                
%=\frac{1}{2} \sum_q e_q^2 q^-/k'^-\;\Delta q^N(x_p). ~~~.
%\label{eq:iqpm2}                                                                
%\end{equation}                                                                  
                                                                                
The use of the improved scaling variable can also have              
a non-trivial consequence on                                                    
the Ellis-Jaffe sum rule violation reported by                                  
several experimental groups.  
For the sake of simplicity,                       
we consider only  valence                                                    
quark contributions to $g^N_1(\nu,Q^{2})$ and                                   
neglect contributions                                                       
from the sea. 
In the SU(6) quark-spectator model \cite{DQ,DQ2} the quark helicity
distributions for the valence quarks read
\begin{equation}
\begin{array}{clcr}
\Delta u_{v}(x,{\bf k}_{\perp})
=u_{v}^{\uparrow}(x,{\bf k}_{\perp})-u_{v}^{\downarrow}(x,{\bf k}_{\perp})
\\=
 -\frac{1}{18}a_V(x,{\bf k}_{\perp})W_V(x,{\bf k}_{\perp})
+\frac{1}{2}a_S(x,{\bf k}_{\perp})W_S(x,{\bf k}_{\perp});\\
\Delta d_{v}(x,{\bf k}_{\perp})
=d_{v}^{\uparrow}(x,{\bf k}_{\perp})-d_{v}^{\downarrow}(x,{\bf k}_{\perp})
=-\frac{1}{9}a_V(x,{\bf k}_{\perp})W_V(x,{\bf k}_{\perp}),
\label{eq:sfdud}
\end{array}
\end{equation}
where $W_D(x,{\bf k}_{\perp})$ is the correction factor 
due the Wigner rotation effect \cite{Ma91}
\begin{equation}
W_D(x,{\bf k}_{\perp})=\frac{(k^+ +m)^2-{\bf k}^2_{\perp}}
{(k^+ +m)^2+{\bf k}^2_{\perp}} 
\end{equation}
with $k^+=x {\cal M}$ and 
${\cal M}=\frac{m^2_q+{\bf k}^2_{\perp}}{x}
+\frac{m^2_D+{\bf k}^2_{\perp}}{1-x}$.
Thus we  can write the Ellis-Jaffe and Bjorken sums
as
\begin{equation}
\begin{array}{clcr}
S^p_{EJ}=\frac{1}{9}<W_S>-\frac{1}{54}<W_V>;
\\
S^n_{EJ}=\frac{1}{36}<W_S>-\frac{1}{36}<W_V>,
\label{eq:SumEL}
\end{array}
\end{equation}
and
\begin{equation}
\begin{array}{clcr}
S_{B}=S^p_{EJ}-S^n_{EJ}=\frac{1}{12}<W_S>+\frac{1}{108}<W_V>,
\label{eq:SumB}
\end{array}
\end{equation}
where
\begin{equation}
<W_D>=
\int_{0}^{1} [{\rm d} x_B] [{\rm d}^2 {\bf k}_\perp]
{\cal K} \; %\frac{q^-}{k'^-} 
%q^-/k'^-\; 
a_D(x_p) W_D(x_p).
%\;dx_B.            
\end{equation}
 
We calculate the Ellis-Jaffe and Bjorken sums with
the above adopted parameters and find for set I:\\
$S^p_{EJ}=0.210$, $S^n_{EJ}=-0.001$, and $S_B=0.209$ at $Q^2=3$ GeV$^2$
and 
$S^p_{EJ}=0.214$, $S^n_{EJ}=-0.0003$, and $S_B=0.214$ at $Q^2=10$ GeV$^2$,\\
which is far off the experimental values
$S_B(E143)=0.163\pm 0.010 \pm 0.016$ and 
$S_B(SMC)=0.199\pm 0.038$.
For set II one gets closer to the  data:\\
$S^p_{EJ}=0.174$, $S^n_{EJ}=-0.012$, and $S_B=0.186$ at $Q^2=3$ GeV$^2$
and
$S^p_{EJ}=0.186$, $S^n_{EJ}=-0.011$, and $S_B=0.214$ at $Q^2=10$ GeV$^2$,\\
which is in good agreement with the experimental values of the Bjorken
sum but still far off the Ellis-Jaffe sums:
$S^p_{EJ}(E143)=0.127\pm 0.004 \pm 0.010$,
$S^n_{EJ}(E143)=-0.037\pm 0.008 \pm 0.011$ at $<Q^2>=3$ GeV$^2$
and 
$S^p_{EJ}(SMC)=0.136\pm 0.011 \pm 0.011$,
$S^n_{EJ}(SMC)=-0.063\pm 0.024 \pm 0.013$ at $<Q^2>=10$ GeV$^2$.
Let us note for comparison that the naive Ellis-Jaffe sum rule would 
result in $S^p_{EJ}(SMC)=0.171\pm 0.006$, see ref. [6]. Obviously 
the sea quarks, which are  missing in the BHL wave function, are important for
the Ellis-Jaffe sum rule. 
Work on an extension of the BHL wave functions to include 
intrinsic sea quarks is 
on the way \cite{Bro96} and it will be very interesting to use the resulting
wave functions to calculate the Ellis-Jaffe sum rule. This will, however, still
require substantial work.\\  
Fig.~1 shows the
spin asymmetries                                                 
$A_1^p(x)=2xg_1^p(x)/F_2^p(x)$ and $A_1^n(x)=2xg_1^n(x)/F_2^n(x)$               
resulting for the parameters of sets I and II. It is obvious that both sets
do not fit the data well.
The discrepancies occure mainly for small $x$ where one would expect 
the sea quark contribution to be sizeable.
 
In summary, we analysed the Gottfried, Bjorken, and Ellis-Jaffe sums 
in an     
improved quark-parton model description of nucleon structure functions.           
It is found that the effects                                                     
from finite quark masses and transversal motion                                         
are noticeable but not large, in contrast to claims in the literature. 
Furthermore we found that a constraint derived from the size of the 
higher twist corrections to the Bjorken 
and Ellis-Jaffe sum rules substantially
reduce the allowed parameter range, which is important for all 
applications of this model. 
The same observation is probably true for other model wave-functions.
This fact high-lights the far reaching 
importance of a precise determination of higher twist effects.
The observed discrepancies for the Gottfried and 
Ellis-Jaffe 
sum rules are reduced but not explained by our kinematic corrections. 
A drawback of the specific BHL wave functions we used is their lack of 
an intrinsic sea quark distribution. We are currently working on an 
improvement of these wave functions  \cite{Bro96}.
                                                                               
\noindent                                                                       
{\large \bf ACKNOWLEDGMENTS}                                                    
                                                                                
We would like to thank S.~J.~Brodsky for useful discussions
during this work.
This work was                                                              
supported in part by the                                                        
Alexander von Humboldt Foundation, the Chinese National Science                 
Foundation grant No.~19445004 and Academia Sinica grant No.~94KJZ063.           
A.S. acknowledges support from DFG, BMBF and MPI f\"ur Kernphysik.

\break                                                                          
\noindent                                                                       
{\large \bf Figure Captions}                                                    
\renewcommand{\theenumi}{\ Fig.~\arabic{enumi}}                                 
\begin{enumerate}                                                               
\item                                                                           
The spin asymmetries $A_1^p(x)$ and $A_1^n(x)$ calculated in the                  
light-cone SU(6) quark-spectator model \cite{DQ2} with
the Martin-Roberts-Stirling ($S'_0$) parametrizations of unpolarized
quark distributions \cite{Mar93}.                                              
The data
are EMC($\bigtriangleup$), SMC($\Box$), 
and E143($\bigcirc$) for $A_1^p(x)$   
and E142($\Diamond$)  for $A_1^n(x)$ \cite{EMC,SMC,E142,E143}.
%The $\bullet$, $\diamond$, and                           
%$\circ$ data                                                                    
%are the EMC \cite{EMC}, SMC \cite{SMC} and E143 \cite{E143} 
%$A_1^p(x)$   
%and $\Box$ the E142 \cite{E142} $A_1^n(x)$ \cite{SMC,E142}            
%results respectively.                                                           
The solid and dashed curves are the results                                               
for parameter set I and II for  $A_1^p(x)$ and $A_1^n(x)$ 
calculated at $Q^2=5$ GeV$^2$. 
%Obviously there is a remaining discrepancy.
\end{enumerate}                                                                 
                                                                                

\newpage                                                                        
                                                                                
                                                                                
\begin{thebibliography}{99}                                                     
                                                                                
\bibitem{GSR}   K.~Gottfried,                                                   
                Phys.~Rev.~Lett. {\bf 18} (1967) 1174.                          
                                                                                
\bibitem{NMC91} NM Collab., P.~Amaudruz {\it et al.},                           
                Phys.~Rev.~Lett. {\bf 66} (1991) 2712;
                M.~Arneodo {\it et al.}, 
                Phys.~Rev.~{\bf D 50} (1994) R1.                             
                                                                                
\bibitem{Ma93}  See, e.g., B.~-Q.~Ma, A.~Sch\"afer, and W.~Greiner,             
                Phys.~Rev.~{\bf D 47} (1993) 51,                                
                and references therein.                                         
                                                                                
\bibitem{EJSR}                                                                  
                J.~Ellis and R.~L.~Jaffe,                                       
                Phys.~Rev.~{\bf D 9} (1974) 1444;                               
                           {\bf 10} (1974) 1669(E).                             
                See, also, M.~Gourdin,                                          
                Nucl.~Phys.~{\bf B 38} (1972) 418.                              
                                                                                
\bibitem{EMC} 
                EM Collab., J.~Ashman {\it et al.}, 
                Phys.~Lett.~{\bf B 206} (1988) 364;
                Nucl.~Phys.~{\bf B 328} (1989) 1.

\bibitem{SMC} 
                SM Collab., B.~Adeva {\it et al.}, 
                Phys.~Lett.~{\bf B 302} (1993) 533;
                {\bf B 357} (1995) 248;    
                D.~Adams {\it et al.}, {\it ibid.}
                {\bf B 329} (1994) 399; {\bf B 339} (1994) 332(E).

\bibitem{E142}
                E142 Collab., P.~L.~Anthony {\it et al.}, 
                Phys.~Rev.~Lett.~{\bf 71} (1993) 959.

\bibitem{E143}
                E143 Collab., P.~L.~Anthony {\it et al.},
                Phys.~Rev.~Lett.~{\bf 74} (1995) 346; 
                K.~Abe {\it et al.}, {\it ibid.} {\bf 75}
                (1995) 25. 

\bibitem{Ans94}                                                                 
                For a recent review, see, e.g., M.~Anselmino,                   
                A.~Efremov, and E.~Leader, 
%                CERN-TH/7216/94,                     
                Phys.~Rep. {\bf 261} (1995) 1.                                         
                                                                                
\bibitem{Ma86}  B.~-Q.~Ma,                                                      
                Phys.~Lett.~{\bf B 176} (1986) 179;                             
                                                                                
                B.~-Q.~Ma and J.~Sun,                                           
                Int.~J.~Mod.~Phys.~{\bf A 6} (1991) 345,                        
                in this paper the kinematical factor $q^-/(q^-+k^-)$            
                in Eq.~(2.12) should be ${\cal K}=q^-/k'^-$ 
                (or $q^-/(k'^- -k^+ P_c^-/P_c^+)$).                             
                                                                                
\bibitem{LC}    S.~J.~Brodsky, in {\it Lectures on Lepton Nucleon               
                Scattering and  Quantum Chromodynamics},                        
                edited by A.~Jaffe and D.~Ruelle                                
                (Birkh\"auser, Boston, 1982), p.~255;                           
                                                                                
                S.~J.~Brodsky, T.~Huang, and G.~P.~Lepage,                      
                in {\it Particles and Fields}, edited by A.~Z.~Capri            
                and A.~N.~Kamal (Plenum, New York, 1983), p.~143.               
                                                                                
\bibitem{Ma91}  B.-Q.~Ma, J.~Phys.~{\bf G 17} (1991) L53;                       
                                                                                
                B.-Q.~Ma and Q.-R.~Zhang, Z.~Phys.~{\bf C 58}                   
                (1993) 479;                                                     
                                 
                S.~J.~Brodsky and F.~Schlumpf,                       
                Phys.~Lett.~{\bf B 329} (1994) 111.                             

\bibitem{Ros79} D.~A.~Ross and C.~T.~Sachrajda, Nucl.~Phys.~{\bf B              
                149} (1979) 497.                                                
                                                                                
\bibitem{Pre91} G.~Preparata, P.~G.~Ratcliffe, and J.~Soffer,                   
                Phys.~Rev.~Lett. {\bf 66} (1991) 687.                           
                                                                                
\bibitem{Pi}    E.~M.~Henley and G.~A.~Miller,                                  
                Phys.~Lett.~{\bf B 251} (1990) 453;                             
                                                                                
%\bibitem{Kum91a}                                                               
                S.~Kumano, Phys.~Rev.~{\bf D 43} (1991) 59;                     
                {\bf D 43} (1991) 3067;                                         
                                                                                
%\bibitem{Sig91}                                                                
                A.~Signal, A.~W.~Schreiber, and A.~W.~Thomas,                   
                Mod.~Phys.~Lett.~{\bf A 6} (1991) 271.                          
                                                                                
\bibitem{Ma92}  B.~-Q.~Ma, Phys.~Lett.~{\bf B 274} (1992) 111.                  
                                                                                
\bibitem{Mar90} A.~D.~Martin, W.~J.~Stirling, and R.~G.~Roberts,                
                Phys.~Lett.~{\bf B 252} (1990) 653;                             
                                                                                
                S.~D.~Ellis and W.~J.~Stirling,                                 
                Phys.~Lett.~{\bf B 256} (1991) 258.                             
                                                                                
\bibitem{Kap91} L.~P.~Kaptari and A.~Yu.~Umnikov,                               
                Phys.~Lett.~{\bf B 272} (1991) 359.                             
                                                                                
\bibitem{Epe92} L.~N.~Epele, H.~Fanchiotti, C.~A.~Carc\'ia Canal,               
                and R.~Sassot,                                                  
                Phys.~Lett.~{\bf B 275} (1992) 155.                             
                                                                                
\bibitem{Saw93}                                                                 
                M.~Sawicki and J.~Vary, Phys.~Rev.~Lett.~{\bf 71}               
                (1993) 1320.                                                    
 
\bibitem{DQ}    See, e.g., R.~Carlitz and J.~Kaur,                              
                Phys.~Rev.~Lett.~{\bf 38} (1977) 673;                           
                                                                                
                A.~Sch\"afer, Phys.~Lett.~{\bf B 208} (1988) 175.

\bibitem{DQ2}
                B.-Q.~Ma, preprint BIHEP-TH-95-30, hep-ph/9604423, 
                Phys.~Lett.~{\bf B} (1996) in press.               

\bibitem{LCQM}  See, {\it e.g.}, 
                H.~J.~Weber, Ann.~Phys.~(N.Y.) {\bf 207} (1991) 417;

                B.~-Q.~Ma, Z.~Phys.~{\bf A 345} (1993) 321;
         
                F.~Schlumpf, Phys.~Rev.~{\bf D 48} (1993) 4478;    

                T.~Huang, B.~-Q.~Ma, and Q.~-X.~Shen,
                Phys.~Rev.~{\bf D 49} (1994) 1490.
 
\bibitem{Stein} E.~Stein, P.~G\'ornicki, L.~Mankiewicz, A.~Sch\"afer, 
                and W.~Greiner, Phys. Lett. {\bf B343} (1995) 369;

                E.~Stein, P.~G\'ornicki, L.~Mankiewicz, and A.~Sch\"afer,
                Phys. Lett. {\bf B353} (1995) 107;
               
                L.~Mankiewicz, E.~Stein, and A.~Sch\"afer,
                Preprint hep-ph/9510418.

\bibitem{Web94}                                                                                
                H.~J.~Weber, 
%                Phys.~Lett.~{\bf B 287} (1992)14;
                Phys.~Rev.~{\bf D 49} (1994) 3160.                 
 
\bibitem{Mar93} A.~D.~Martin, W.~J.~Stirling, and R.~G.~Roberts,                
                Phys.~Lett.~{\bf B 306} (1993) 145.
%                Phys.~Rev.~{\bf D 47} (1993) 867.

\bibitem{Bro96}
               S.~J.~Brodsky and B.-Q.~Ma, preprint SLAC-PUB-7126
               and BIHEP-TH-96-13, hep-ph/9604393, Phys.Lett.B (1996) in press. 
\end{thebibliography}
\end{document}